\documentstyle[psfig]{l-aa}
\newcommand{\mo}    {\hbox{${\rm M}_{\odot}$}}

\newcommand{\kms}   {\hbox{${\rm km\,s}^{-1}$}}

\newcommand{\cmsq}  {\hbox{${\rm cm}^{-2}$}}
\newcommand{\cmcb}  {\hbox{${\rm cm}^{-3}$}}

\newcommand{\htwo}  {\hbox{${\rm H}_2$}}                
\newcommand{\jeno}  {$J$=1$\leftarrow$0}                
\newcommand{\jowt}  {$J$=2$\leftarrow$1}                
\newcommand{\ffam}{\hbox{$\,.\!\!^{\prime}$}}

\begin{document}
\topmargin=1.0cm
\thesaurus{ 09.01.1;  
	    09.07.1;  
	    09.13.2;  
	    11.05.1;  
            11.09.1;  
	    11.09.4;  
	         }
   \title{Molecular absorption and its time variations in Cen A}
%
   \author{T.~Wiklind\inst{1}, F.~Combes\inst{2}}
   \offprints{T.~Wiklind, tommy@oso.chalmers.se}
   \institute{Onsala Space Observatory, S--43992 Onsala, Sweden
   \and
              DEMIRM, Observatoire de Paris, 61 Av. de l'Observatoire,
              F--75014 Paris, France
             }
   \date{Received date; Accepted date}
   \maketitle
   \markboth{Molecular absorption in Cen A}
   {Molecular absorption in Cen A}
%
\begin{abstract}

New high quality absorption spectra of \jeno\ HCO$^+$, HCN, HNC
and the \jowt\ line of CS towards the center of the radio core
of Cen A (NGC\,5128) are presented.  In addition the absorption
profile of the \jeno\ line of H$^{13}$CO$^+$ has been detected for
the first time, revealing that most absorption features have 
low opacities. 
The new HCO$^+$ spectrum allows a comparison with results obtained
more than 7 years earlier. No significant change in the spectrum
is found. From this remarkable result, constraints can be put on
the minimum size of the radio source in the millimeter range
($>$500 AU).
A comparison of abundance ratios between different absorption
components suggests that the absorbing gas is essentially low
density gas, with both low excitation and kinetic temperatures.
Relative molecular abundances are compatible with those of the 
Milky Way.
A parametrization of the high signal--to--noise spectrum of HCO$^+$
is presented for future reference when looking for spectral changes.

\end{abstract}
\section{Introduction}

The nearby giant elliptical galaxy Cen A (NGC\,5128) contains a warped
disk of dust and dense gas. Optically the disk is seen as obscuration
against a background of the old stellar population belonging to the
elliptical galaxy. The presence of H$\alpha$ emission in the disk
(cf. Nicholson et al. 1992)
suggest that formation of massive stars are currently taking place
in the disk.

The disk is also a source of molecular line emission. Several studies
have shown that it contains about $3 \times 10^{8}$\,\mo\ of \htwo.
The distribution of the molecular gas has been traced by the CO emission
(Eckart et al. 1990a, Quillen et al. 1992, Rydbeck et al. 1993, Wild et
al. 1997). The emission extends to a galactocentric distance of
approximately 1\,kpc.
The molecular gas distribution and its kinematics is consistent with a
thin disk which is severly warped (Quillen et al. 1992). The main components
are a ring or spiral arm at a galactocentric distance of $\sim$800\,pc\ (adopting
a distance of 3\,Mpc to Cen A, which means that 1'' corresponds to
14.5\,pc.) and a circumnuclear ring at a radius of $\sim$100\,pc (Israel
et al. 1990, Rydbeck et al. 1993). The inner ring is seen as high velocity
wings in spectra towards the center when the angular resolution is better
than 25--30''. It has also been imaged with the aid of deconvolution of single
dish CO(2--1) data (Rydbeck et al. 1993). The inner ring is inclined relative
to the outer disk but aligned perpendicular to the inner radio jet. 
The molecular gas properties of the disk appears to be similar to those
found in normal spiral galaxies (cf. Israel et al. 1990, Eckart et al. 1990,
Wild et al. 1997).

The radio core is hidden behind a large column of dense obscuring gas.
The combination of a strong radio continuum source and a large column
of dense gas makes the line of sight towards the center of Cen A a rich
source of molecular absorption lines.

The properties of the gas seen in absorption is still largely unknown.
Several studies have come up with conflicting results, both concerning
the location of the absorption components relative to the nucleus as well
as the temperature and density of the gas.
The HI absorption towards the nucleus shows three absorption components;
one strong at the systemic velocity around 552\,\kms\ and two redshifted
ones at 596 and 609\,\kms, respectively. Towards the inner jet, only the
main absorption at 552\,\kms\ is seen (van der Hulst et al. 1983).
This has been interpreted
as evidence that the main 552\,\kms\ line is situated far out in the
disk, while the redshifted lines are situated very close to the nucleus,
possibly falling in to the center.
However, Seaquist \& Bell (1990) report a detection of redshifted H$_2$CO
$\lambda$2cm absorption against the inner jet at a velocity of
$\sim$576\,\kms.
This is not necessarily a proof against the redshifted component being
situated close to the nucleus. The inner jet is seen at a projected
distance of $\sim$20''
from the core, which corresponds to $\sim$300\,pc. The inner molecular
disk can be extended on these scales (cf. Israel et al. 1991, Rydbeck
et al. 1993, Hawarden et al. 1993). The molecular absorption lines seen
in the millimeter range only occurs towards the radio core of Cen A,
since the inner jet has a steep radio
spectrum with a completely negligible continuum flux at mm wavelengths.

Molecular absorption lines seen in our Galaxy (cf. Lucas \& Liszt 1996)
and towards high redshift galaxies (Wiklind \& Combes 1996ab, 1995, Combes
\& Wiklind 1996) almost exclusively arise in very cold gas
(in terms of excitation temperature). Although the abundance
ratio of HCN/HNC imply that the kinetic temperature can be in the range
10--20\,K, the excitation temperature is comparable to the cosmic
microwave background.
This suggests diffuse gas with n(H$_2$) $< 10^{3}$ cm$^{-3}$.
Are the molecular absorption lines seen in Cen A likewise coming from
diffuse gas? Unfortunately very few multiline transitions of the same
molecule have been observed. One exception is H$_2$CO, for which
Seaquist \& Bell (1990) derive an upper limit to the excitation
temperature of 3.9\,K. Several OH absorption features have been 
detected by van Langevelde et al. (1995), their interpretation is
complicated by the presence of maser lines. Some come from diffuse
gas, and some features point to dense clumps
(n(H$_2$) $>$ 10$^4$ cm$^{-3}$).
The three lowest rotational lines of CO have been seen in absorption,
both the lines around the systemic velocity and the redshifted
components. The mere detection of the CO(3--2) line (cf. Israel et
al. 1991) implies that the excitation temperature is relatively high
(10--20\,K). For CO, however, the analysis is complicated by
confusion with emission, especially for the two lowest transitions.

In this paper we present new high quality observations of the
HCO$^+$(1--0), HCN(1--0), HNC(1--0) and CS(2--1) absorption lines,
as well as the previously unobserved H$^{13}$CO$^+$(1--0) transition,
in order to shed some light on the physical properties and location
of the molecular absorption towards the radio core of Cen A.

In Sect.\,2 we present the observations, in Sect.\,3 we identify
the different absorption components and question their possible
time variations and in Sect.\,4 we derive column densities and
abundance ratios. In Sect.\,5 we discuss the implied properties
of the absorbing gas in Cen A, and its assumed distance from the center.

\begin{figure*}
\psfig{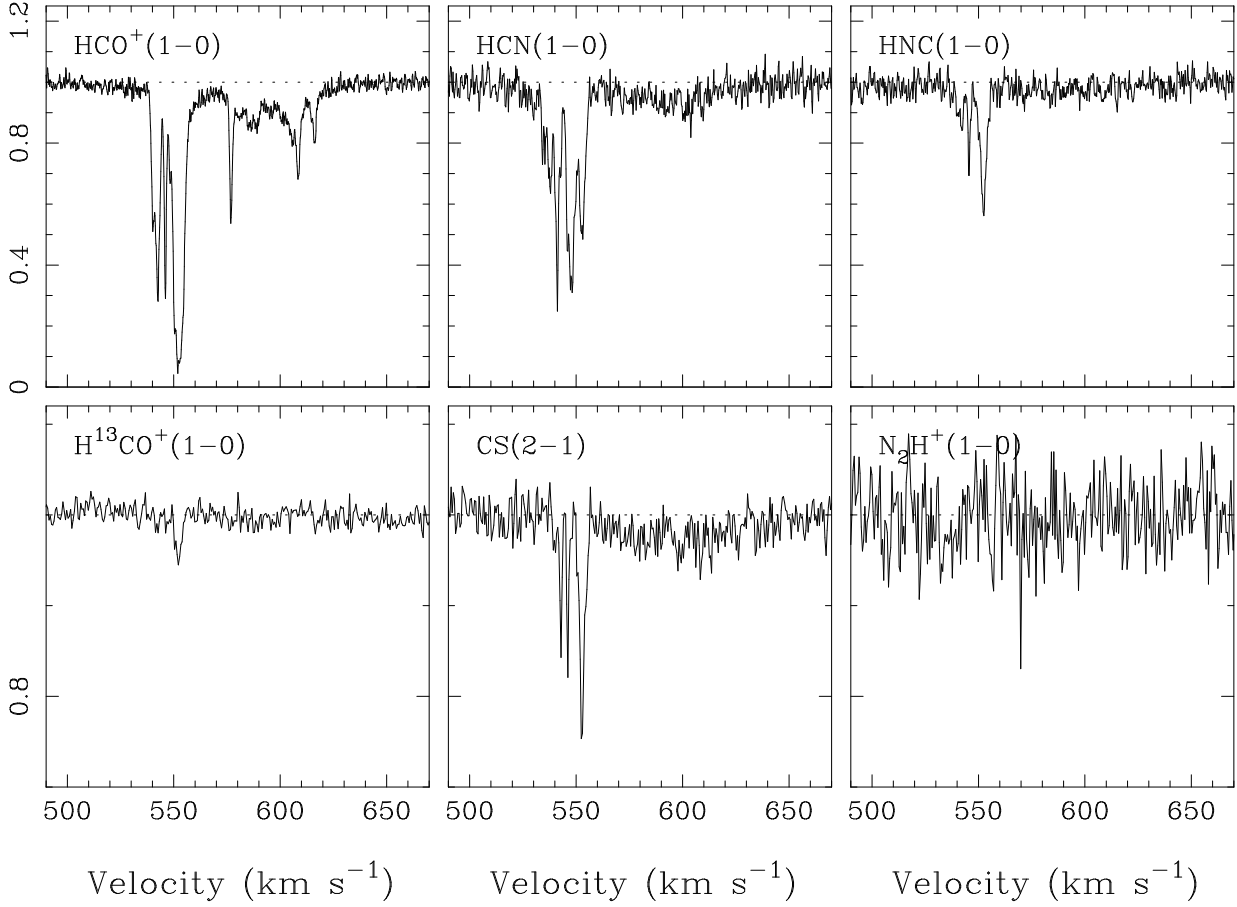}
\caption[]{Spectra of the observed transitions: HCO$^+$(1--0),
H$^{13}$CO$^+$(1--0), HCN(1--0), HNC(1--0), CS(2--1) and
N$_2$H$^+$(1--0). Only N$_2$H$^+$ remains undetected.
The spectra have been normalized to a continuum level of unity.
The velocity scale is heliocentric and the frequency definition
is relativistic (see text).
The velocity resolution is 0.2\,\kms\ for
HCO$^+$, 0.4\,\kms\ for HCN, HNC and CS, and 0.6\,\kms\ for H$^{13}$CO$^+$
and N$_2$H$^+$. }
\end{figure*}

\begin{table*}
\begin{flushleft}
\caption[]{Observed molecules}
\small
\begin{tabular}{l|cccc}
\hline
 & & & & \\
\multicolumn{1}{c|}{Molecule}                    &
\multicolumn{1}{c}{Transition}                   &
\multicolumn{1}{c}{Frequency$^{a)}$}             &
\multicolumn{1}{c}{$T_{\rm cont}$}               &
\multicolumn{1}{c}{Observing date}               \\
\multicolumn{1}{c|}{ }                           &
\multicolumn{1}{c}{J$\longrightarrow$J--1}       &
\multicolumn{1}{c}{GHz}                          &
\multicolumn{1}{c}{mK$^{b)}$}                    &
\multicolumn{1}{c}{ }                            \\
 & & & & \\
\hline
 & & & & \\
H$^{13}$CO$^+$ & 1--0 & 86.754294 & 409,355,308 & Dec95, Jul96, Aug96 \\
HCN            & 1--0 & 88.630415 & 320 & Jul96                       \\
HCO$^+$        & 1--0 & 89.188518 & 410,340 & Dec95, Jul96            \\
HNC            & 1--0 & 90.663450 & 325 & Jul96                       \\
N$_2$H$^+$     & 1--0 & 93.173500 & 314 & Dec95                       \\
CS             & 2--1 & 97.980968 & 350 & Jul96                       \\
 & & & & \\
\hline
\end{tabular}
\ \\
a)\ The rest--frequency of the observed molecules ($\nu_0$ in Eqs\,1--2)
taken from Lovas (1992). \\
b)\ In the $T_{\rm A}^{*}$ temperature scale. Conversion to Jansky:
25\,Jy/K.
\end{flushleft}
\end{table*}

\begin{figure*}
\psfig{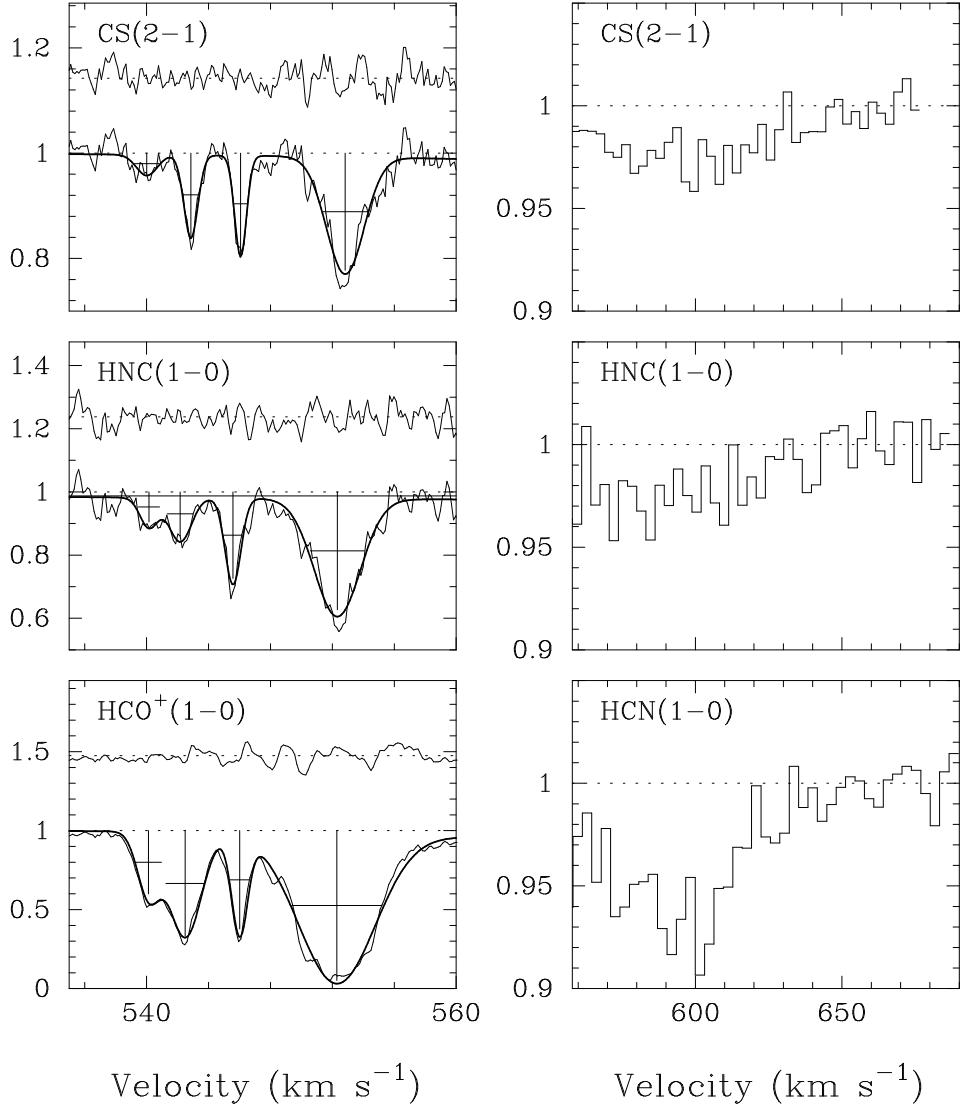}
\caption[]{The four main lines in the Low Velocity complex is shown to the left
for CS(2--1), HNC(1--0) and HCO$^+$(1--0). Also shown are the results from
a 5--component gaussfit. The residuals shown above the lines, are spectral
values minus the fitted values. The fift component corresponds to the High
Velocity complex, shown to the right. The HV complex has been binned to a
velocity resolution of 3\,\kms. Notice the different scales.
These five components define the main absorption lines (see Table\,2).}
\end{figure*}

\section{Observations}

The observations were done with the 15--m Swedish--ESO Submillimeterwave
Telescope (SEST) at La Silla in Chile. The HCO$^+$, H$^{13}$CO$^+$ and
N$_2$H$^+$ lines were observed in December 1995, the HCN, HNC and CS
lines in July 1996, where we also obtained some additional observations
of the HCO$^+$ and H$^{13}$CO$^+$ lines. A final observing session in
August 1996 only involved the H$^{13}$CO$^+$ line.

For all the observations we used the 3--mm SIS mixer, giving a system
temperature in the range 140--160\,K at the observed frequencies (see
Table\,1). The low $T_{\rm sys}$ allowed us to obtain spectra with
significantly better signal--to--noise than previous observations of
the molecular absorption line system in Cen A. As backend we used
the high resolution Acoustical Spectrometer (AOS), giving a resolution of
43\,kHz or 0.14\,\kms. Due to a Lorentzian response of
the AOS, the actual velocity resolution of the spectra are slightly worse
than this. The effective resolution is $\sim$0.2\,\kms.
The observations were done in a dual beamswitch mode, with a beamthrow
of $\pm$11\ffam9 in azimuth at a switchfrequency of 6\,Hz.
Both the receiver system and the weather were good and stable during
the observations. The pointing was checked on the continuum of Cen A
itself and the rms variations were less than 3''. The good pointing
accuracy is evident as small rms variations of the continuum level
in 10 minute averages of spectra obtained during 8 hours of observations.

The observed continuum levels at the different frequencies are given in
Table\,1.
Some of the high resolution spectra show a distinct curvature due to the
presence of emission. We subtracted a second order baseline from the average
spectra of each molecular line. Redshifted absorption features in the
velocity range 560--640\,\kms\ can be seen in the HCO$^+$, HCN, HNC
and CS spectra. The region between 490--660\,\kms\ was therefore
omitted from the fitting procedure.
The bandwidth of the spectrometer does not cover the full extent of
the emission seen in low resolution data (cf. Israel et al. 1992). 
We derived the continuum level at the edges of the spectra in order
to eliminate contribution from the emission. For HCO$^+$ and HCN,
however, the presence of emission produces a flux $\la$2\% too high. 
This transforms into an error in the optical depth calculations which
is 10\% for the deepest HCO$^+$ absorption. For a line with a
depth of 0.5 relative to the continuum, the error is $<$3\%.

The 3--mm continuum flux decreased by $\sim$20\% between December
1995 and July 1996. This is consistent with intrinsic variations
of the millimetric flux of Cen A (cf. Tornikoski et al. 1996).
The dispersion in the continuum flux determinations are caused by
pointing and calibration errors. The latter is of the order 10\%,
whereas the former is likely to be less than this. The opacities
of the absorption lines are, however, independent on the
pointing\footnote{This is true as long as the only contribution
to changes in the observed continuum level is due to pointing
offsets.}, given only by the continuum to line ratio. In the
following we have normalized all the continuum levels to  unity
using the fluxes given in Table\,1.
The continuum measured at $\lambda$3mm does get a contribution
from the far-infrared dust emission in the disk. This is, however,
insignificant. With a dust temperature of 36\,K, as derived from
the IRAS 60 and 100\,$\mu$m fluxes, and a flux of 320.6\,Jy at
100\,$\mu$m, the contribution to the continuum at $\lambda$3mm
is at most 2\%. This is calculated assuming a dust emissivity
$\propto \nu$. A more realistic dust emissivity
($\propto \nu^{\alpha}, \alpha > 1.0$) makes the dust emission
negligble at $\lambda$3mm.

\medskip

The rest--frequencies of the observed molecules are given in Table\,1.
We used $v=550$\,\kms\ as the systemic velocity for Cen A.
All velocities are given as heliocentric with the relativistic
definition of the frequency shift. This means that the center
frequency of the spectrometer is given as
\begin{eqnarray}
\nu_{\rm sky} & =  & \nu_0 \gamma \left(1-\frac{v}{c}\right)\ \ \
{\rm Relativistic}\ ,
\end{eqnarray}
where $v$ is the heliocentric velocity of the source (here 550\,\kms),
$c$ is the speed of light, $\nu_0$ the rest-frequency as given in Table\,1
and $\gamma$ the Lorentz factor $ 1/\sqrt{1-v^2/c^2}$.
The corresponding radio and optical definitions are
\begin{eqnarray}
\nu_{\rm sky} & = & \nu_0 \left(1-\frac{v}{c}\right)\ \ \ \ \ \
{\rm Radio} \nonumber \\
\ \\
\nu_{\rm sky} & = & \nu_0 \left(1+\frac{v}{c}\right)^{-1}\ \ \
{\rm Optical} \ .\nonumber
\end{eqnarray}
The observations in December were in fact done with the LSR velocity
system and using the radio definition of the redshifted frequency.
These observations have been shifted to a heliocentric velocity
system and using the relativistic definition.
In December 1995 $V_{\rm HEL} = V_{\rm LSR} - 2.208$\,\kms.
The velocity difference between the radio and relativistic frequency
definitions amounts to 0.504\,\kms, with the relativistic
definition giving a higher sky-frequency, i.e. the frequency to which
the receiver is in effect tuned to. In order to convert the spectra
obtained with the radio definition to the relativistic definition,
we have to add 0.504\,\kms.
Hence, the data obtained in December 1995 has been shifted by
a total of 2.712\,\kms. Before adding the HCO$^+$ and
H$^{13}$CO$^+$ obtained in December with that obtained in 1996,
we mapped the 1995 spectra into the same channels as the 1996 spectra.

\begin{figure}
\psfig{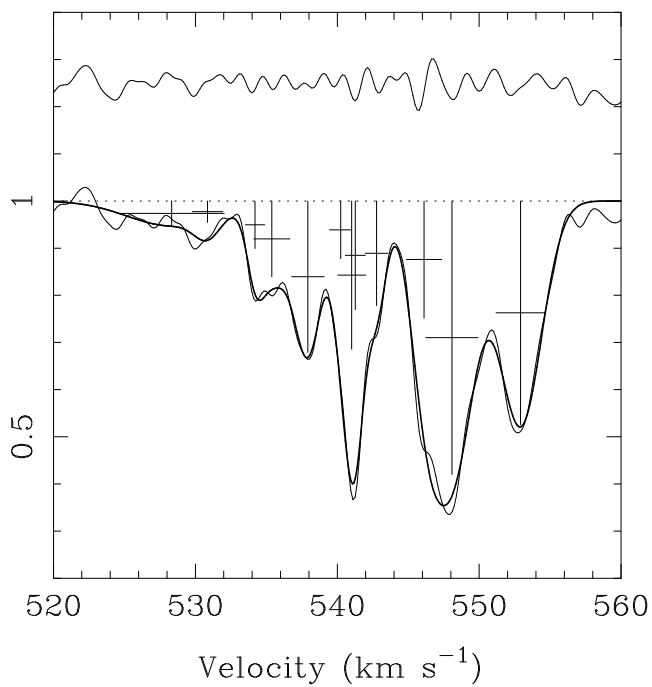}
\caption[]{Decomposition of the hyperfine components of HCN(1--0)
in the LV complex. Four main absorption systems are assumed,
each associated with three hyperfine lines of HCN. The center
velocities of the components have been kept fixed in the fitting
procedure.}
\end{figure}

\begin{table*}
\begin{flushleft}
\caption[]{Main gaussian components}
\small
\begin{tabular}{r|rrr|rrr|rrr|rrr}
\hline
 & & & & & & & & & & & & \\
\multicolumn{1}{c}{ }                            &
\multicolumn{3}{|c|}{HCO$^{+}$(1--0)}            &
\multicolumn{3}{c|}{H$^{13}$CO$^{+}$(1--0)}      &
\multicolumn{3}{c|}{HNC(1--0)}                   &
\multicolumn{3}{c}{CS(2--1)}                     \\
 & & & & & & & & & & & & \\
\hline
 & & & & & & & & & & & & \\
\multicolumn{1}{c}{No.}                          &
\multicolumn{1}{|c}{V$_{0}^{a)}$}                &
\multicolumn{1}{c}{T$_{0}^{b)}$}                 &
\multicolumn{1}{c|}{$\Delta$V$^{c)}$}            &
\multicolumn{1}{c}{V$_{0}^{a)}$}                 &
\multicolumn{1}{c}{T$_{0}^{b)}$}                 &
\multicolumn{1}{c|}{$\Delta$V$^{c)}$}            &
\multicolumn{1}{c}{V$_{0}^{a)}$}                 &
\multicolumn{1}{c}{T$_{0}^{b)}$}                 &
\multicolumn{1}{c|}{$\Delta$V$^{c)}$}            &
\multicolumn{1}{c}{V$_{0}^{a)}$}                 &
\multicolumn{1}{c}{T$_{0}^{b)}$}                 &
\multicolumn{1}{c}{$\Delta$V$^{c)}$}             \\
\multicolumn{1}{c}{ }                            &
\multicolumn{1}{|c}{\kms}                        &
\multicolumn{1}{c}{mK}                           &
\multicolumn{1}{c|}{\kms}                        &
\multicolumn{1}{c}{\kms}                         &
\multicolumn{1}{c}{mK}                           &
\multicolumn{1}{c|}{\kms}                        &
\multicolumn{1}{c}{\kms}                         &
\multicolumn{1}{c}{mK}                           &
\multicolumn{1}{c|}{\kms}                        &
\multicolumn{1}{c}{\kms}                         &
\multicolumn{1}{c}{mK}                           &
\multicolumn{1}{c}{\kms}                         \\
 & & & & & & & & & & & & \\
\hline
 & & & & & & & & & & & & \\
 4 & 540.12 & 399.7 &   1.68 &
            &       &        &
     540.15 &  94.8 &   1.37 &
     540.01 &  39.7 &   1.58 \\
 3 & 542.49 & 668.3 &   2.48 &
            &       &        &
     542.17 & 139.2 &   1.70 &
     542.86 & 158.2 &   1.07 \\
 2 & 546.01 & 622.5 &   1.22 &
            &       &        &
     545.58 & 273.9 &   1.20 &
     546.07 & 192.3 &   0.90 \\
 1 & 552.29 & 948.3 &   5.84 &
     552.16 &  50.1 &   3.03 &
     552.32 & 373.0 &   3.50 &
     552.83 & 221.5 &   2.86 \\
 5 & 593.94 & 139.5 &  50.12 &
            &       &        &
     575.65 &  25.5 & 101.04 &
     596.79 &  28.6 &  62.59 \\
 & & & & & & & & & & & & \\
\hline
\end{tabular}
\ \\
a)\ Heliocentric velocity with relativistic velocity definition
(see text). \\
b)\ Depth of the absorption line measured from the normalised
continuum level
(T$_{\rm A}^{*}$). \\
c)\ Full width at half maximum depth.
\end{flushleft}
\end{table*}

\begin{table*}
\begin{flushleft}
\caption[]{HCN(1--0) hyperfine components for the LV--complex$^{a)}$}
\small
\begin{tabular}{c|rrr|rrr|rrr|cc}
\hline
 & & & & & & & & & & & \\
\multicolumn{1}{c|}{ }                           &
\multicolumn{3}{c|}{F$=1-1$}                     &
\multicolumn{3}{c|}{F$=2-1$}                     &
\multicolumn{3}{c|}{F$=0-1$}                     &
\multicolumn{1}{c}{ }                            &
\multicolumn{1}{c}{ }                            \\
 & & & & & & & & & & & \\
\multicolumn{1}{c|}{Comp.}                       &
\multicolumn{1}{c}{Vel.$^{b)}$}                  &
\multicolumn{1}{c}{T$_{\rm abs}$}                &
\multicolumn{1}{c|}{$\Delta$V$^{c)}$}            &
\multicolumn{1}{c}{Vel.$^{b)}$}                  &
\multicolumn{1}{c}{T$_{\rm abs}$}                &
\multicolumn{1}{c|}{$\Delta$V$^{c)}$}            &
\multicolumn{1}{c}{Vel.$^{b)}$}                  &
\multicolumn{1}{c}{T$_{\rm abs}$}                &
\multicolumn{1}{c|}{$\Delta$V$^{c)}$}            &
\multicolumn{1}{c}{$\int{\tau_{\nu}dV}$}         &
\multicolumn{1}{c}{N$^{d)}$}               \\
\multicolumn{1}{c|}{ }                           &
\multicolumn{1}{c}{km\,s$^{-1}$}                 &
\multicolumn{1}{c}{mK}                           &
\multicolumn{1}{c|}{km\,s$^{-1}$}                &
\multicolumn{1}{c}{km\,s$^{-1}$}                 &
\multicolumn{1}{c}{mK}                           &
\multicolumn{1}{c|}{km\,s$^{-1}$}                &
\multicolumn{1}{c}{km\,s$^{-1}$}                 &
\multicolumn{1}{c}{mK}                           &
\multicolumn{1}{c|}{km\,s$^{-1}$}                &
\multicolumn{1}{c}{km\,s$^{-1}$}                 &
\multicolumn{1}{c}{cm$^{-2}$}                    \\
 & & & & & & & & & \\
\hline
 & & & & & & & & & \\
1 & 552.92 & 474 & 3.48 & 548.08 & 580 & 3.68 & 541.01 & 315  & 1.96  &
5.99 & $2.65 \times 10^{13}$ \\
2 & 546.11 & 248 & 2.48 & 541.27 & 231 & 1.42 & 534.20 & 101  & 1.37  &
1.26 & $5.56 \times 10^{12}$ \\
3 & 542.77 & 222 & 1.62 & 537.93 & 321 & 1.96 & 530.86 &  45: & 2.16: &
1.29 & $5.70 \times 10^{12}$ \\
4 & 540.24 & 122 & 1.57 & 535.39 & 161 & 2.53 & 528.33 &  $<$55 & 7.3: &
1.10 & $4.89 \times 10^{12}$ \\
 & & & & & & & & & & \\
\hline
\end{tabular}
\ \\
a)\ The absorption complex around the systemic velocity 553\,\kms. \\
b)\ Heliocentric velocities with relativistic definition (see text). \\
c)\ Full width at half intensity. \\
d)\ Derived with an excitation temperature $T_{\rm x} = 5$\,K.
\end{flushleft}
\end{table*}

\begin{figure*}
\psfig{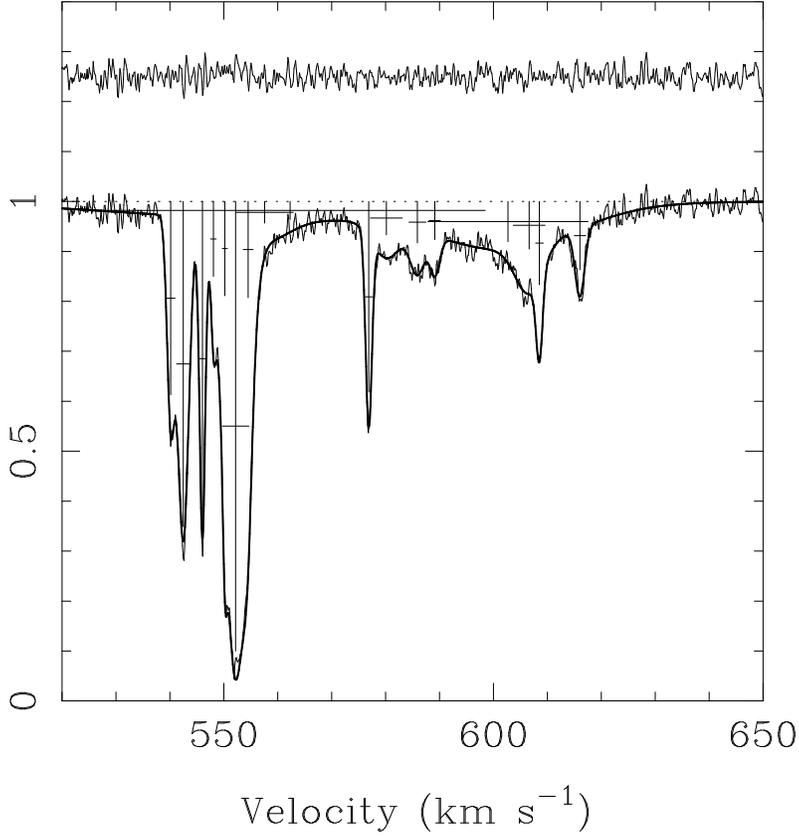}
\caption[]{Parametrization of the HCO$^+$(1--0) spectrum
through a fit of 17 gaussian components. Data for the
gaussian components are given in Table.\,5. The residual
is spectral value minus fitted value and has an rms similar
to that of the original spectrum away from the absorption
complexes. The velocity resolution is 0.2\,\kms.}
\end{figure*}

\begin{figure*}
\psfig{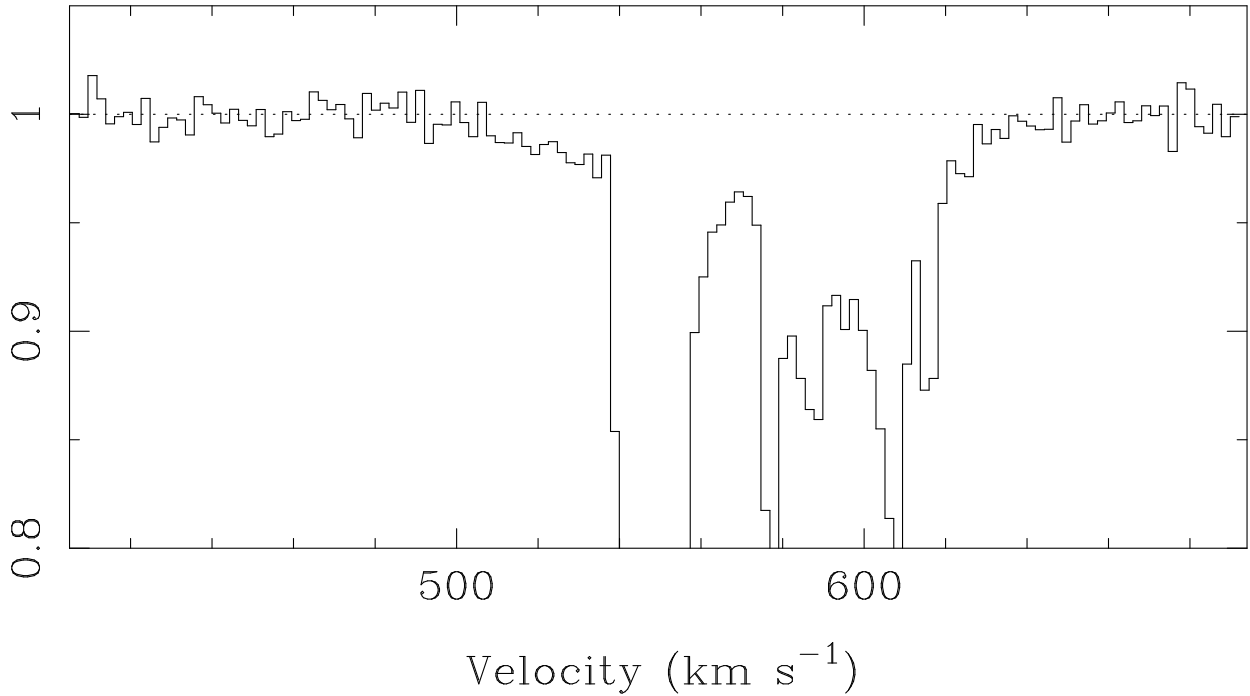}
\caption[]{The full extent of the HCO$^+$(1--0) spectrum.
The data has been binned to give a velocity resolution of
2.1\,\kms. The blueshifted absorption `wing' extends to
$\sim$500\,\kms, while the absorption extends to
$\sim$640\,\kms\ on the redshifted side (see also Fig.\,4).}
\end{figure*}

\section{Results}

\subsection{The spectra}

The absorption spectra are shown in Fig.\,1, with the continuum
normalized to unity.
The N$_2$H$^+$(1--0) line is not detected, despite the presence of
an apparent line seen in Fig.\,1. None of the 7 hyperfine components
of this transition correspond to this line, moreover, its appearance
differs from the lines detected in the other molecules. The N$_2$H$^+$
`line' is an artefact caused by the receiver at this particular
frequency. For HCO$^+$, HCN, HNC and CS we can identify two major
absorption complexes.

\begin{itemize}

\item The strongest absorption feature is situated close to
the systemic velocity of 552\,\kms, with a few relatively
strong blue shifted absorption lines. We will henceforth call
this complex of absorption lines the Low Velocity complex,
or the LV complex. This complex consists of at least 4 different
absorption components, extending between 540--556\,\kms. The
high quality HCO$^+$ spectrum shows evidence for 7 components.

\item Extended absorption at velocities redshifted relative
to the systemic velocity is most evident in the HCO$^+$ line,
but can be seen in the CS, HCN and HNC lines as well. We will
refer to this complex as the High Velocity complex or the HV
complex. The HV complex consists of a broad `diffuse' absorption,
extending from about 560\,\kms\ to approximately 640\,\kms.
The HCO$^+$ spectrum also shows several narrow absorption
lines superposed on the broad absorption. These lines are absent
from the other lines. The HV complex is not clearly evident
in the H$^{13}$CO$^+$ spectrum shown in Fig.\,1, but when binning
the spectrum the HV complex is present at significant level of
several sigma.

\end{itemize}

For reference we identify 5 main absorption lines, visible
in all but the H$^{13}$CO$^+$ and N$_2$H$^+$ spectrum. No.
1--4 defines the LV complex and no. 5 is the broad diffuse
HV complex.
In Fig.\,2 we show the LV and HV complexes in more detail.
Superposed on the LV complex are the results of a 5--component
gauss fit; four components for the LV complex and one for
the HV complex (not shown). The fit parameters are given in
Table\,2. The residuals are similar to the rms noise level
of the spectra, except for HCO$^+$, where the high
signal--to--noise reveals the several additional
unresolved components in the LV complex (of course,
the HV complex of HCO$^+$ is poorly fitted by the fifth
gauss component due to the presence of several narrow absorption
lines). We will return to the HCO$^+$ spectrum in Sect.\,3.3.

\begin{table}
\begin{flushleft}
\caption[]{Ratios of HCN(1--0) hyperfine components for the
LV--complex$^{a)}$}
\small
\begin{tabular}{c|cc|cc}
\hline
 & & & & \\
\multicolumn{1}{c|}{ }                           &
\multicolumn{2}{c|}{R$_{12}$}                    &
\multicolumn{2}{c}{R$_{01}$}                     \\
 & & & & \\
\multicolumn{1}{c|}{Comp.}                       &
\multicolumn{1}{c}{Peak$^{b)}$}                  &
\multicolumn{1}{c|}{Integrated$^{c)}$}           &
\multicolumn{1}{c}{Peak$^{b)}$}                  &
\multicolumn{1}{c}{Integrated$^{c)}$}            \\
 & & & & \\
\hline
 & & & & \\
1 & 0.8 & 0.8 & 0.5 & 0.3 \\
2 & 1.1 & 1.9 & 0.4 & 0.4 \\
3 & 0.7 & 0.6 & 0.1 & 0.2 \\
4 & 0.8 & 0.5 & $<$0.3 & $<$1.0 \\
 & & & & \\
\hline
\end{tabular}
\ \\
a)\ The absorption complex around the systemic velocity
553\,\kms. \\
b)\ Ratio of the peak anteanna temperatures. \\
c)\ Ratio of the component integrated intensity. \\
\end{flushleft}
\end{table}

\begin{table}
\begin{flushleft}
\caption[]{HCO$^+$(1--0) gaussian components}
\small
\begin{tabular}{rrrr}
\hline
\multicolumn{4}{c}{HCO$^{+}(1-0)$}               \\
\hline
 & & & \\
\multicolumn{1}{c}{No.}                          &
\multicolumn{1}{c}{V$_{0}^{a)}$}                 &
\multicolumn{1}{c}{T$_0$}                        &
\multicolumn{1}{c}{$\Delta$V$^{b)}$}             \\
\multicolumn{1}{c}{ }                            &
\multicolumn{1}{c}{\kms}                         &
\multicolumn{1}{c}{mK}                           &
\multicolumn{1}{c}{\kms}                         \\
 & & & \\
\hline
 & & & \\
 1 & 540.14 & 373.3 &  1.68 \\
 2 & 542.48 & 639.1 &  2.48 \\
 3 & 546.03 & 580.4 &  1.33 \\
 4 & 548.07 & 131.3 &  1.12 \\
 5 & 550.19 & 163.4 &  1.03 \\
 6 & 552.15 & 854.6 &  4.85 \\
 7 & 554.50 & 199.5 &  1.95 \\
 8 & 556.01 &  60.4 & 11.19 \\
 9 & 567.18 &  36.7 & 80.67 \\
10 & 576.89 & 359.6 &  1.47 \\
11 & 580.16 &  68.6 &  6.28 \\
12 & 585.92 &  83.7 &  3.18 \\
13 & 589.12 &  77.4 &  2.18 \\
14 & 602.52 &  76.9 & 26.92 \\
15 & 606.68 &  94.7 &  5.83 \\
16 & 608.52 & 157.1 &  1.47 \\
17 & 616.02 & 133.9 &  2.21 \\
 & & & \\
\hline
\end{tabular}
\ \\
a)\ Heliocentric velocity with relativistic velocity definition
(see text). \\
b)\ Full width at half minimum.
\end{flushleft}
\end{table}

\subsection{The hyperfine components of HCN}

HCN is not part of Table\,2 since each absorption system here
consists of three hyperfine transitions. The high quality
HCN(1--0) spectrum allows us to make a decomposition of the
hyperfine lines by keeping the center velocities for each
absorption system fixed at the value derived from the other
molecules and only varying the intensity and width of the
lines. The fitting is complicated by a near overlap of the
F$=$0--1 hyperfine line of component 1 with the F$=$2--1
line of component 2. The fitting is not perfect and the results
actually suggests the presence of at least a fifth component
(adding three more hyperfine lines). However, with the present
signal--to--noise ratio and overlapping components, we use 4
absorption lines and smooth the HCN spectrum before fitting.
The resulting decomposition is shown in Fig.\,3 and the
hyperfine components of the 4 main absorption lines in the
LV complex is given in Table\,3. The results are robust for
all the F$=$1--0 lines and for all the F$=$2--1 lines,
except for absorption line no. 2 which is situated within
0.3\,\kms\ from the F$=$0--1 line of absorption line no. 1.
The F$=$0--1 lines of line no. 3 and 4 are too weak to
give completely reliable results. Nevertheless, the hyperfine
line ratios, $R_{12}= I_{1-0}/I_{2-1}$ and
$R_{02}= I_{0-1}/I_{2-1}$, as derived from the gaussian
decomposition are given in Table\,4.

\subsection{Parametrization of the HCO$^+$(1--0) spectrum}

In order to facilitate future studies of the small scale
structure of the molecular ISM along the line of sight towards
the radio core in Cen A, we have parametrized the high
quality HCO$^+$(1--0) absorption line shown in Fig.\,1.
The parametrization was done by fitting a number of gauss
components to the spectrum. By starting out with a relatively
small number of components and increasing the number until
the residual does not show any large deviations from the
rms noise of the spectrum, we found that a minimum of 17
components were needed. The reality of some of thes
components should not be taken too seriously, since some
absorption features may not be resolved and therefore do not
show real gaussian profiles. The gaussian components can,
however, be used for a precise representation of the
absorption spectrum obtained by us and used for comparison
with spectra obtained with different telescopes, spectrometers
and spectral resolutions.

The gaussian components are given in Table\,5 as the peak
depth (measured from the normalized continuum), the full
width at half intensity and the center velocity. The
identification of these components are given from 1 to 17,
and should not be confused by the identification of the
5 main absorption lines as given in Table\,2. The
HCO$^+$(1--0) spectrum with the fitted curve overlaid
is shown in Fig.\,4, together with a residual spectrum.

\medskip

In addition to the HV and LV complexes there is an extended
blueshifted absorption `wing' in the HCO$^+$ spectrum. This
is evident already in Fig.\,4, but is more clearly seen in
a smoothed spectrum. In Fig.\,5 we show the HCO$^+$ spectrum
binned to a velocity resolution of 2.1\,\kms.
The blueshifted wing extends to about 500\,\kms.
On the redshifted part of the spectrum, the broad absorption
extends out to about 640\,\kms.
In Fig.\,6 we show the original data before baseline subtraction.
The second order baseline which is subsequently used in the
subtraction is shown as a dashed line. The peak of the baseline
falls at $\sim$550\,\kms, as it should, despite the small and
uneven regions where it was fitted.
The depression at 500--540\,\kms\ can be seen quite clearly. It
is also present in the low resolution data presented by
Israel (1992).
Hence, the molecular absorption along the line of sight to the
radio core in Cen A occurs continuously over 140\,\kms.
It is not possible to determine if the absorption in the range
500--540\,\kms\ is a single blueshifted feature or if it is
associated with the red `wing' at $\sim$615--640\,\kms.

\subsection{Variations in the absorption lines?}

The radio source in Cen A has a classical steep--spectrum
double lobe structure, a steep--spectrum inner jet and a
flat--spectrum core. VLBI observations at 8.4\,GHz indicate
that the core has a very small extent $\la$2\,milliarcseconds
(mas) (Jones et al. 1996). The core is self absorbed at
2.3\,GHz and brightens up to frequencies of $\sim$20\,GHz.
This small core is the likely source of the continuum
emission seen at millimeter wavelengths. At a distance of
3\,Mpc the extent of the core is thus only $\la$0.03\,pc,
or 6000\,AU. The core may even be considerably smaller;
the variability at millimeter wavelengths observed by
Kellerman (1974) suggests a size of the order a light day,
or 175\,AU. In addition to the core, a narrow
steep--spectrum jet is conspicuous at 8.4\,GHz
(Jones et al. 1996); its intensity is already much weaker
than the core, and will be completely negligible at mm
wavelengths.

Small scale structure in the molecular ISM has been
directly observed down to $\sim$2000\,AU (e.g. Wilson
\& Walmsley 1989, Falgarone et al. 1992), while VLBI
techniques allow detection of structures of 25\,AU in
the HI absorption medium in front of 3C radio sources
(Diamond et al. 1989). Multi--epoch observations of
21cm absorption against high velocity pulsars also allowed
detection of opacity variations of the ISM on a range of
scales from 5\,AU to 100\,AU (Frail et al. 1994). In
the molecular ISM, scales of the order of 10\,AU have
been inferred, through the time variations of H$_2$CO
absorption lines (Marscher et al. 1993).
Moore \& Marscher (1995) confirmed these H$_2$CO absorption
variations over a few years, in front of several point radio
sources (3C111, NRAO150 and BL Lac). Through numerical
simulations Marscher \& Stone (1994) were able to derive
constraints on the fractal structure of molecular clouds,
from the time variability detections. The mean number of
small clumps along the line of sight should be larger than
previously thought, i.e. the size spectrum of clumps should
be a steeper power--law, constraining the fractal dimension.

The surprising fact in these time variations of absorption
features is that the involved gas appears to be diffuse.
In the case of the 21cm absorption against pulsars, where
small--scale opacity structures are detected towards {\it all}
line of sights, the mean opacities are between 0.1 and 2.5,
corresponding to N(HI) as low as 10$^{19}-10^{20}$\,cm$^{-2}$
(Frail et al. 1994). Also in the case of molecular 6cm H$_2$CO
absorptions, the mean optical depth is very low (Moore
\& Marscher 1995). Of course variations of opacities are
much more easy to detect when the opacity is low, since
saturated profiles ($\tau >> 1$) are acutely sensitive to
small variations in the noise level. But it was not expected
to find contrasted small scale structures, and clumps, in such
a diffuse gas. The opacity variations dectected are
quite high (half of them have $\delta\tau > 0.1$, Frail et al.
1994), so they cannot be accounted for by mild density
fluctuations. Clumps with density larger than
10$^5-10^6$ cm$^{-3}$ are implied (Moore \& Marscher 1995).
These clumps could represent 10--20\% of the total column density.

\begin{figure}
\psfig{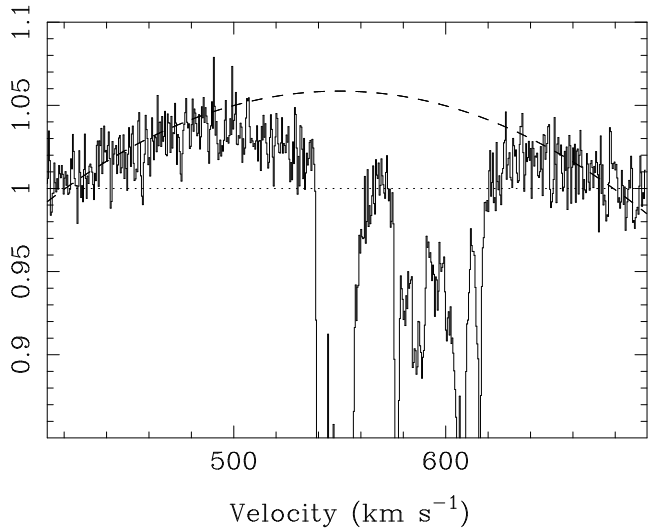}
\caption[]{The HCO$^+$(1--0) spectrum at a velocity resolution of
0.7\,\kms\ before baseline removal. A second order baseline fit is
shown as a dashed line. The blue `wing' at 500--540\,\kms\ is clearly
visible.}
\end{figure}

\medskip

The same physical conditions appear to be true for the molecular
absorbing gas in front of the Cen A radio source. The opacities
and column densities must be low in average, since the isotopic
H$^{13}$CO$^+$ line is observed with about the expected abundance
ratio with respect to the HCO$^+$ line. In these conditions,
opacity variations should be easy to detect.
The gaseous disk in Cen A has a rotational velocity of more
than 250\,\kms. The value is somewhat uncertain due to the
strong warp of the disk and could be higher (cf. van Gorkom et
al. 1990, Rydbeck et al. 1993).
A rotational velocity of 250\,\kms\ corresponds to a shift of
the gas along the line of sight towards the radio core in Cen A
by $\sim$50\,AU per year.

The first molecular absorption line observations in Cen A was
obtained in 1976 (Gardner \& Whiteoak 1976), of formaldehyde.
However, it wasn't until 1989 that the first high resolution
molecular absorption line data was obtained, which can be used
for a detailed comparison with the present data (cf. Eckart et
al. 1990). Hence, the time span is about 7 years and a comparison
with this data is thus sensitive to spatial scales in the ISM of
300--500\,AU.

The HCO$^+$(1--0) spectrum published by Eckart et al. (1990)
does not have as good signal--to--noise ratio as the current
one and any comparison is limited by the rms noise in the
1989 data. In Fig.\,7 we show our HCO$^+$(1--0) spectrum and
the one obtained by Eckart et al. (1990) in 1989. Both spectra
have been binned to a velocity resolution of 0.6\,\kms. The
lower panel in Fig.\,7 shows the difference between the two
spectra; $1995/96-1989$ with equal weighting.
A small residual feature can be seen in the difference
spectrum at a velocity of 552\,\kms, with a depth
$\sim$10\% of the normalized continuum level.
This feature corresponds to the deepest absorption
line. It could, however, be an effect of a slightly incorrect
continuum flux in the 1989 data. If the continuum flux
has been overestimated, the line--to--continuum ratio
decreases (and the difference will show up for the strongest
absorption feature).
Likewise, an incorrect continuum level for
the 1995/96 data could mimic the residual feature. The
latter is not likely, since the observing conditions
for the 1995/96 data were very good.
Also, a comparison between the HCO$^+$ data obtained
in December 1995 and July 1996 do not show
any residual (Fig.\,8). Although the time span is only
six months, the considerably higher quality of the data
makes a comparison sensitive to very small changes in
both the location of the absorption lines and their depths.
The sensitivity for shifts in velocity is very good and the
limits are less than $-0.2$ to $+0.3$\,\kms; The difference
depends on the shape of the absorption profiles.
This allows us to put an upper limit of 10\% to changes in
the HCO$^+$(1--0) absorption line in Cen A. If the molecular
ISM in Cen A have small scale structures similar to those
found in our Galaxy, this negative result implies that the
background continuum source at millimeter wavelengths has
an extent $\ga$500\,AU. The limit given by VLBI is 6000\,AU.
However, it should be kept in mind that if the absorbing
gas is made up of a large number of small clouds ($>$200 per
absorption feature) with angular extent smaller than
that of the continuum source, changes will be difficult
to detect since the clumps would tend to average out over
time (e.g. Marscher \& Stone 1994). Only future high--quality
data can solve the question regarding variations in the
absorbing gas in Cen A

\begin{figure}
\psfig{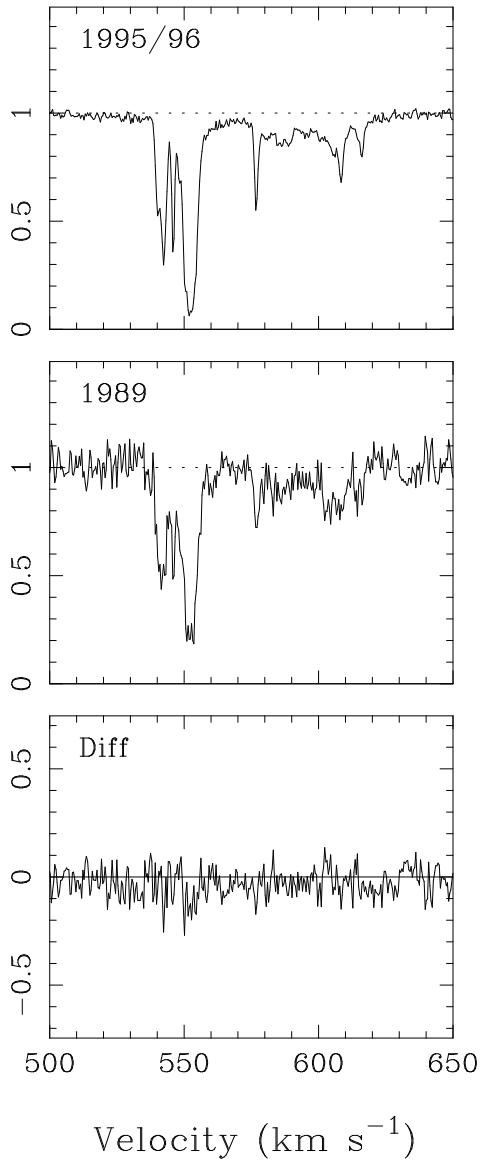}
\caption[]{HCO$^+$(1--0) spectra obtained in 1995/96 (present paper)
and in 1989 (Eckart et al. 1990). The difference ($1995/96-1989$)
is shown in the third panel. The spectra have been binned to a velocity
resolution of 0.6\,\kms\ before subtraction. No significant differences
can be found.}
\end{figure}

\begin{table*}
\begin{flushleft}
\caption[]{Column densities and abundance ratios for the main
absorption components$^{a)}$.}
\scriptsize
\begin{tabular}{l|rrrr|c|c}
\hline
 & & & & & & \\
\multicolumn{1}{c|}{ }                           &
\multicolumn{4}{c|}{Low Velocity complex$^{b)}$} &
\multicolumn{1}{c|}{Average LV$^{c)}$}           &
\multicolumn{1}{c}{High Velocity complex}        \\
 & & & & & & \\
\multicolumn{1}{c|}{Molecule}                    &
\multicolumn{1}{c}{1}                            &
\multicolumn{1}{c}{2}                            &
\multicolumn{1}{c}{3}                            &
\multicolumn{1}{c|}{4}                           &
\multicolumn{1}{c|}{1--4}                        &
\multicolumn{1}{c}{5}                            \\
 & & & & & & \\
\multicolumn{1}{c|}{ }                           &
\multicolumn{1}{c}{552.3 km\,s$^{-1}$}           &
\multicolumn{1}{c}{545.9 km\,s$^{-1}$}           &
\multicolumn{1}{c}{542.5 km\,s$^{-1}$}           &
\multicolumn{1}{c|}{540.0 km\,s$^{-1}$}          &
\multicolumn{1}{c|}{538--560 km\,s$^{-1}$}       &
\multicolumn{1}{c}{600.0 km\,s$^{-1}$}           \\
 & & & & & & \\
\hline
 & & & & & & \\
HCO$^{+}$        & 13.659 & 12.595    & 12.959    & 12.490    &
13.790 & 13.420 \\
H$^{13}$CO$^{+}$ & 11.809 & $<$10.615 & $<$10.777 & $<$10.657 &
11.916 & 11.821 \\
HCN              & 13.424 & 12.745    & 12.756    & 12.689    &
13.683 & 13.187 \\
HNC              & 12.883 & 12.280    & 12.150    & 11.851    &
13.099 & 12.851 \\
CS               & 13.022 & 12.439    & 12.415    & 12.016    &
13.246 & 13.355 \\
 & & & & & & \\
\hline
 & & & & & & \\
$\frac{{\rm HCO}^+}{{\rm H}^{13}{\rm CO}^{+}}$&71&$>$96&$>$152&
$>$64&75&$>$40\\
 & & & & & & \\
$\frac{{\rm HCO}^+}{{\rm HCN}}$                & 1.7 & 0.7   &
1.6    & 0.6   & 1.3 & 1.7 \\
 & & & & & & \\
$\frac{{\rm HCO}^+}{{\rm HNC}}$                & 6.0 & 2.1   &
6.4    & 4.4   & 3.8 & 3.7 \\
 & & & & & & \\
$\frac{{\rm HCO}^+}{{\rm CS}}$                 & 4.3 & 1.4   &
3.5    & 3.0   & 3.5 & 1.2 \\
 & & & & & & \\
$\frac{{\rm HCN}}{{\rm HNC}}$                  & 3.5 & 2.9   &
4.0    & 6.9   & 3.8 & 2.2 \\
 & & & & & & \\
$\frac{{\rm HCN}}{{\rm CS}}$                   & 2.5 & 2.0   &
2.2    & 4.7   & 2.7 & 0.7 \\
 & & & & & & \\
$\frac{{\rm HNC}}{{\rm CS}}$                   & 0.7 & 0.7   &
0.5    & 0.7   & 0.7 & 0.3 \\
 & & & & & & \\
\hline
\end{tabular}
\ \\
a)\ An excitation temperature of 5\,K has been used for all
calculations. \\
b)\ The velocities are the average center velocities from
all the molecules. Column densites are given as
$\log(N/{\rm cm}^{-2})$.\\
c)\ For HCN the LV complex is defined by the velocity
intervall 525--560\,km\,s$^{-1}$ due to the hyperfine
components. \\
\end{flushleft}
\end{table*}

\section{Column densities and abundance ratios}

\subsection{Basic assumptions}

In order to derive accurate column densities from the absorption
lines, we have to make three assumptions. These are
\begin{enumerate}

\item assume a value for the covering factor $f$ of the absorbing
molecular gas with respect to the extent of the radio core in
Cen A.

\item assume a value for the excitation temperature $T_{\rm x}$
of the absorbing gas.

\item assume that each molecular species is in local thermodynamical
equilibrium, i.e. that all the rotational levels are characterized
by the same excitation temperature (called then the rotational
temperature $T_{\rm rot}$).

\end{enumerate}

\medskip

The covering factor $f$ is unknown for Cen A, but due to the
small size of the radio core ($\la$2.0\,mas$=$6000\,AU), it
is likely to be close to unity for the molecular gas. Also,
the extinction towards the core of Cen A at optical wavelengths
has been estimated to be $A_{\rm V}=15 \pm 5$\,mag (Eckart
et al. 1990), corresponding to a substantial amount of obscuring
material. The depth of the deepest HCO$^+$(1-0) absorption
line (no. 1 in Table\,2) is 0.95 when the continuum level is
unity. If the line is saturated, this means that 95\% of the
continuum source is covered by molecular gas. If, on the other
hand, the absorption line is not saturated, the covering factor
is larger. The abundance ratio of HCO$^+$/H$^{13}$CO$^+$ for
the deepest absorption line is $\sim$70. This ratio would be
lower than the real $^{12}C$/$^{13}C$ ratio if
(1) the HCO$^+$ line was saturated and (2) if isotopic
fractionation augments the H$^{13}$CO$^+$ abundance. The high
HCO$^+$/H$^{13}$CO$^+$ ratio therefore shows that the HCO$^+$
line is unlikely to be strongly saturated. The nondetection of
N$_2$H$^+$ is consistent with this scenario.
The observed optical depth is nevertheless 3.0 for the deepest
absorption component.
The other absorption components are not as deep, but the upper
limits to the HCO$^+$/H$^{13}$CO$^+$ ratio at the corresponding
velocities (Table\,7) implies that these lines are not saturated
and, hence, that their covering factors are always larger than
their depth in the normalized spectra.
In the following we will assume that $f = 1$.

\medskip

\begin{figure}
\psfig{figure=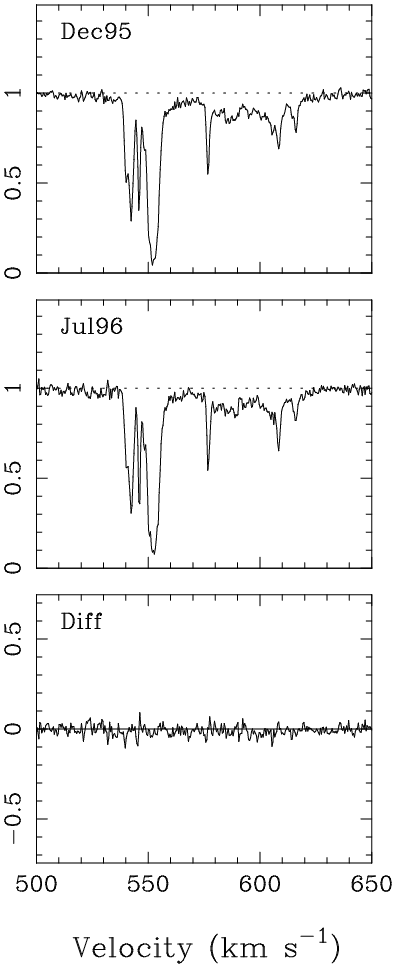,bbllx=60mm,bblly=40mm,bburx=150mm,bbury=230mm,width=8.5cm,angle=0}
\caption[]{HCO$^+$(1--0) spectra obtained in December 1995
and in July 1996. The difference is shown in the third panel.
The spectra have been binned to a velocity resolution of 0.6\,\kms\ before
subtraction. No differences can be found.}
\end{figure}

The excitation temperature defines the relative population
of two levels and can be derived if the velocity integrated
optical depths of two rotational transitions of the same
molecule can be determined. This is not the case for Cen A.
Furthermore, in order to derive the total column density we
must link the fractional level population to all the levels.
This is done by invoking a weak LTE--approximation and
assuming that $T_{\rm x} = T_{\rm rot}$, where $T_{\rm rot}$
is a temperature which governs the fractional population of
all rotational levels in a given molecule. The LTE
approximation is weak in the sense that it does not imply
that $T_{\rm rot}$ equals the kinetic temperature and it
allows for different molecules to have different
$T_{\rm rot}$. With this approximation we can use the
partition function $Q(T_{\rm x})$ to express the total
column density $N_{\rm tot}$ as
\begin{equation}
N_{\rm tot} = {8\pi \over c^{3}} {\nu^{3} \over g_{J} 
A_{J,J+1}} f(T_{\rm x}) \ \int \tau_{\nu} dV \ ,
\end{equation}
where $\int \tau_{\nu} dV$ is the observed optical depth
integrated over the line for a given transition, $g_J$
the statistical weight for rotational level $J$, $A_{J,J+1}$
the Einstein radiative transition coefficient for
levels $J$ and $J+1$, and
\begin{equation}
f(T_{\rm x}) = {Q(T_{\rm x}) \exp(E_{J}/kT_{\rm x}) \over
1-\exp(-h\nu/kT_{\rm x})}.
\end{equation}
The excitation temperature is most likely low for the gas
seen in absorption. The reason is that a high $T_{\rm x}$
quickly depopulates the lower rotational levels and
decreases their opacity. For
$T_{\rm x} \ga 10$\,K, $\int \tau_{\nu} dV \propto N/T_{\rm x}^2$.
Hence, along a line of sight with a mixture of molecular
gas components of similar column density but with different
excitation temperatures, absorption lines of ground
transitions will preferentially sample the excitationally
coldest gas. Excitationally cold gas does not necessarily
imply that the kinetic temperature is low. HCO$^+$, HCN
and HNC are thermalized at \htwo\ densities of $10^5$\,\cmcb.
A molecular gas with densities lower than this would
therefore give these molecules a low excitation temperature.
In a survey of HCO$^+$(1--0) absorption and emission in our
Galaxy, Lucas \& Liszt (1996) find only one case out of
eighteen where emission is associated with the absorption.
This shows that the excitation temperature in the gas sampled
through absorption is indeed very low.
Also denser molecular gas seen in absorption has a low
excitation temperature. Greaves \& Williams (1992)
measured $T_{\rm x}$ using CS(2--1) and CS(3--2) for
several clouds towards Sgr\,B2 and found $T_{\rm x}$
to be always $<$4\,K.

The detection of a relatively strong absorption of
CS(2--1) does not necessarily imply either a high
$T_{\rm x}$ or a high density. The J$=$1 level of
CS has an energy corresponding to 2.35\,K and should
be significantly populated by the cosmic microwave
background radiation. We searched for the CS(3--2) line
and came up with a very weak detection of the main line
at 552\,\kms\ (line no. 1 in Table\,2). The line is too
weak to allow a determination of the excitation temperature,
but the mere difficulty in detecting this line is a strong
indication that the excitation temperature is low. The
J$=$2 level of CS has an energy corresponding to 7.05\,K.
The CO molecule is detected in both the J$=$1--0,
J$=$2--1 and J$=$3--2 transitions. The energy
J$=$1 and J$=$2 levels correspond to temperatures
of 5.53 and 16.6\,K, respectively. The HV complex is
tentatively detected in CO J$=$3--2 (Israel et al. 1991)
\footnote{The HV is not visible in CO(1--0) with single
dish telescopes, because of confusion with larger-scale
emission. Interferometer CO(1--0) data obtained with the
JCMT and CSO telescopes on Mauna Kea resolve out the
emission component and the HV absorption components are
obvious (JCMT Newsletter No. X).}.
CO has an electric dipole moment more than $10^3$ times lower
than either HCO$^+$, HCN, HNC and CS, and is thermalized at
\htwo\ densities which are $10^2$ times lower. Hence, the
excitation temperature of CO is likely to be higher. This
is consistent with the weak LTE assumption made above, where
$T_{\rm x}$ can vary between different molecular species.

Since we have observed molecules with a large electric
dipole moment, we will use $T_{\rm x}=5$\,K for all
the lines when deriving column densities. If the excitation
temperature is close to the cosmic microwave background
temperature, we overestimate the column densities of HCO$^+$,
HCN and HNC by a factor 2.0. For CS the factor is 1.5. If,
on the other hand, the excitation temperature is 10\,K, we
would underestimate the HCO$^+$, HCN and HNC column densities
by a factor 3.1 and the CS by a factor 2.5.

A low $T_{\rm x}$ means that only the lowest rotational
levels are significantly populated and that the assumption
that one single temperature governs the overall population
distribution, i.e. $T_{\rm x}=T_{\rm rot}$ is likely
to be valid. The abundance ratios are very insensitive
to the assumed excitation temperature, only depending on
the weak LTE assumption. Hence, the column densities should
be accurate to within a factor of 2, while the abundance
ratios are very robust estimates.

\begin{figure*}
\psfig{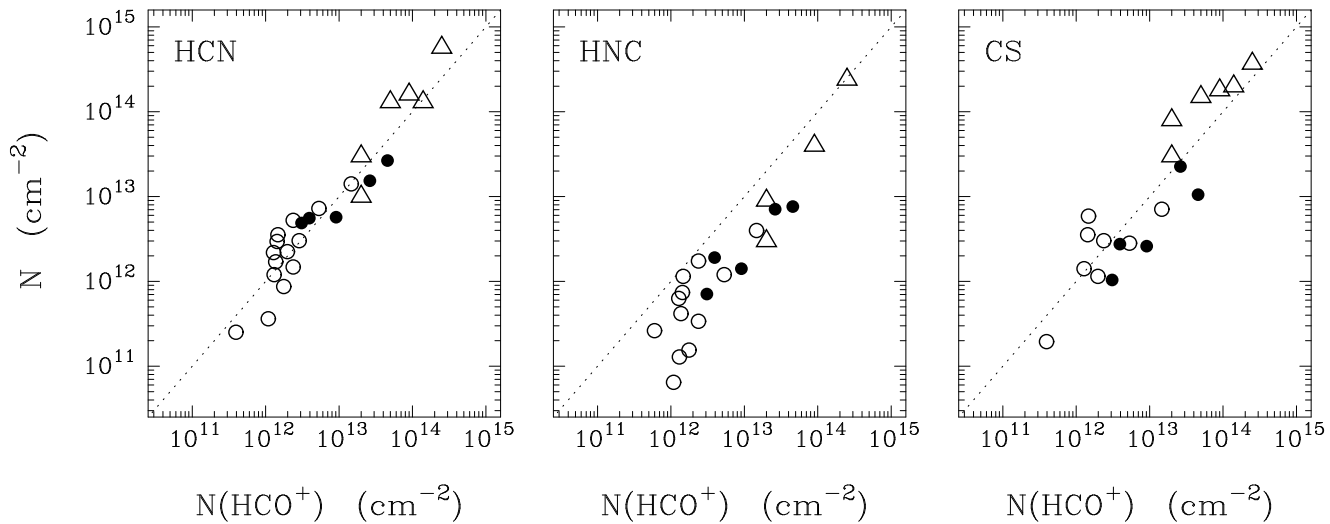}
\caption[]{HCO$^+$ column density vs. HCN, HNC and CS column density.
Filled circles represent our data on Cen A, open circles
data from Galactic diffuse clouds (Lucas \& Liszt 1993, 1994,
private communication)
and triangles data towards SgrB2 from Greaves \& Nyman (1996).
The HCO$^+$ column densities of Lucas \& Liszt have been multiplied by
a factor 1.4 due to different use of the electric dipole
moment. The dotted line represents a one--to--one correspondance
between the column densities.}
\end{figure*}

\begin{table}
\begin{flushleft}
\caption[]{Column densities of HCO$^+$ in the High Velocity
absorption complex}
\small
\begin{tabular}{c|cc}
\hline
 & & \\
\multicolumn{1}{c|}{Comp.}                       &
\multicolumn{1}{c}{Velocity$^{a)}$}              &
\multicolumn{1}{c}{$\log{N/{\rm cm}^{-2}}^{b)}$} \\
 & & \\
\multicolumn{1}{c|}{ }                           &
\multicolumn{1}{c}{km\,s$^{-1}$}                 &
\multicolumn{1}{c}{ }                            \\
 & & \\
\hline
 & & \\
5a & 574--580 km\,s$^{-1}$ & 12.621 \\
 & & \\
5b & 583--593 km\,s$^{-1}$ & 12.651 \\
 & & \\
5c & 598--612 km\,s$^{-1}$ & 12.912 \\
 & & \\
5d & 613--620 km\,s$^{-1}$ & 12.410 \\
 & & \\
\hline
\end{tabular}
\ \\
a)\ Velocity region over which the integrated opacity was
obtained. \\
b)An excitation temperature of 5\,K has been used for all
calculations. \\
\end{flushleft}
\end{table}

\subsection{Qualitative results}

We calculated column densities for the four absorption
lines in the LV complex by using the gaussian components
given in Table\,2. For the HV complex we derived the velocity
integrated optical depth directly from the
spectra\footnote{Only the HCO$^+$ spectrum gives values which
are discrepant from the gaussian fits, due to the presence
of several narrow absorption components in the HV complex.}.
The results are presented in Table\,6. For HCN we used the
decomposition of the hyperfine components as given in Table\,3.
For all lines we used an excitation temperature of 5\,K.
Upper limits to the velocity integrated optical depth of
H$^{13}$CO$^+$ was derived as
\begin{equation}
\int \tau_{\nu} dV \le 3 \sigma_{\tau} \sqrt{\Delta V \delta V}\ ,
\end{equation}
where $\sigma_{\tau}$ is the rms of the opacity (0.012),
$\delta V$ the velocity resolution (0.15\,\kms) and $\Delta V$
the assumed line width. For the latter we used 50\% of the
corresponding HCO$+$ line (compare with the line widths of
HCO$^+$ and H$^{13}$CO$^+$ for line no. 1). A 3$\sigma$ upper
limit to the N$_2$H$^+$ column density, not given in Table\,6,
is $5.6 \times 10^{12}$\,\cmsq. This is derived in the same
way as above, but with a $\Delta V$ of 5\,\kms\ in order to
compare it with line no. 1.

In Table\,7 we present column densities and abundance ratios
for 4 clearly defined HCO$^+$(1--0) components in the High Velocity
complex. The components can only be identified in the HCO$^+$
spectrum and are designated 5a--d.
They are defined in a smoothed spectrum and correspond to
gaussian components 10, 11-13, 14--16 and 17 as given in Table\,5.
Integrated optical depths were derived by integrating over the
velocity intervals given in Table\,7.

\section{Discussion}

\subsection{Type of molecular clouds}

The column density of HCO$^+$ in Cen A ranges between
$3.0 \times 10^{12}$\,\cmsq\  to $4.6 \times 10^{13}$\,\cmsq\ 
(Table\,6). These column densities are all higher than the
onset of CO self--shielding, which occurs at HCO$^+$ column
densities $(1-3) \times 10^{12}$\,\cmsq\ (Lucas \& Liszt 1996).
Hence the absorption in Cen A arises in gas which has a
relatively low C$^+$/C ratio. If the gas in Cen A follows
the same tight correlation between $N(HCO^+)$ and $N(OH)$
as Galactic clouds (Lucas \& Liszt 1996), our HCO$^+$
measurement imply OH column densities in the range
$(0.1-2) \times 10^{15}$\,\cmsq. These $N(OH)$ values are
consistent with those inferred by van Langevelde et al.
(1995) in Cen A.

In Fig.\,9 we plot the column density of HCO$^+$ versus the
column density of HCN, HNC and CS. We also include absorption
line data from Galactic diffuse clouds obtained by Lucas
\& Liszt (1993, 1994, 1996) and somewhat denser clouds seen
towards Sgr\,B2 (Greaves \& Nyman 1996). The latter clouds are
in some cases blended with gas close to the Galactic center.
We have multiplied the HCO$^+$ column densities of Lucas \&
Liszt with a factor 1.4 in order to correct for the use of
different values of the electric dipole moment.
The dotted line in the figure is not a fit to the
data but represents a one--to--one correspondance between
the column densities. The gas in Cen A follows the same
correlation as that of Galactic molecular gas. We have
used an excitation temperature of 5\,K when deriving column
densities, while Lucas \& Liszt and Greaves \& Nyman have
used 2.76\,K. Since the molecules have almost exactly the
same dependence on $T_{\rm x}$, this has no influence on
the result.

Although the excitation temperature is likely to be low
in the molecular gas seen in absorption, the HCN/HNC
ratios implies that the kinetic temperature is rather
high. The formation of HCN and HNC depends on the gas
temperature (cf. Irvine et al. 1987), with HCN preferentially
being formed in warm gas on behalf of its isotopomer HNC.
In the LV complex we find HCN/HNC ratios ranging between
3--7, while the HV complex has a ratio of 2.2. This gives
a kinetic temperature of the LV complex of 20--30\,K, while
the HV gas is characterized by a kinetic temperature of
$\la$10\,K.

A more detailed comparison of the 5 main absorption
components as defined in Table\,2 and 6, shows that
component 1 and 3 (in the LV complex) shows similar
behaviour in their abundance ratios (see Table\,6).
The differences between the components are, however,
relatively small and do not show up as significant
deviations in Fig.\,9. The ratios of the hyperfine
components of HCN (Table\,4) indicate that component
1 and 3 are close to the LTE values;
R$_{12}=0.6$, R$_{02}=0.2$, while component 2
and 4 deviates from LTE. Here we have to keep in mind
that in the decomposition of the hyperfine lines some
components suffer from near overlap and that the
F$=$0--1 line of component 4 is an upper limit.
Nevertheless, it is tantalizing that the two components
with close to LTE ratios also show similar abundance
ratios, while the other two components (plus the HV
complex) differs by factors of 2--3 from each other.
Anomalous excitation of the HCN hyperfine lines is
often seen in Galactic molecular clouds (cf. Guilloteau
\& Baudry 1981, Walmsley et al. 1982), with the R$_{12}$
ratio lower than LTE values in warm clouds and the
R$_{01}$ ratio lower than LTE values in cold clouds.
In dense regions, where the HCN(1--0) line thermalizes,
both ratios tend to unity. This latter case appears to
be the case for component no. 2 in Cen A, at least for
the R$_{12}$ ratio.

In Fig.\,10 we compare the HCO$^+$(1--0) spectrum with the
HI absorption obtained with the VLA (van der Hulst et al.
1983). The HCO$^+$ spectrum has been binned to the same
velocity resolution as the HI data, 6.2\,\kms. A 3--component
gaussian fit to the HCO$^+$ spectrum and the residual is
also shown. The appearance of the spectra agree quite well,
even though the redshifted HCO$^+$ is more spread out in
velocity than the HI. Column densities and  N(HI)/N(HCO$^+$)
ratios are given in Table\,8.
The component at $\sim$580\,\kms\ appears to have a higher
molecular gas fraction than the other two by a factor of
$\sim$3.

In summary, all of the molecular gas components seen in
absorption towards the nucleus of Cen A have a chemistry
similar to that of Galactic diffuse molecular gas. Only
the column densities are higher in Cen A than in typical
Galactic diffuse clouds, which could be due to unresolved
line components (i.e. more clouds) in the line of sight
through the edge--on disk.
The narrow HCO$^+$ lines seen in the HV complex reveal a
relatively diffuse gas with lower abundances and a low
kinetic temperature (T$_{\rm k} \le$ 10K).

\begin{figure}
\psfig{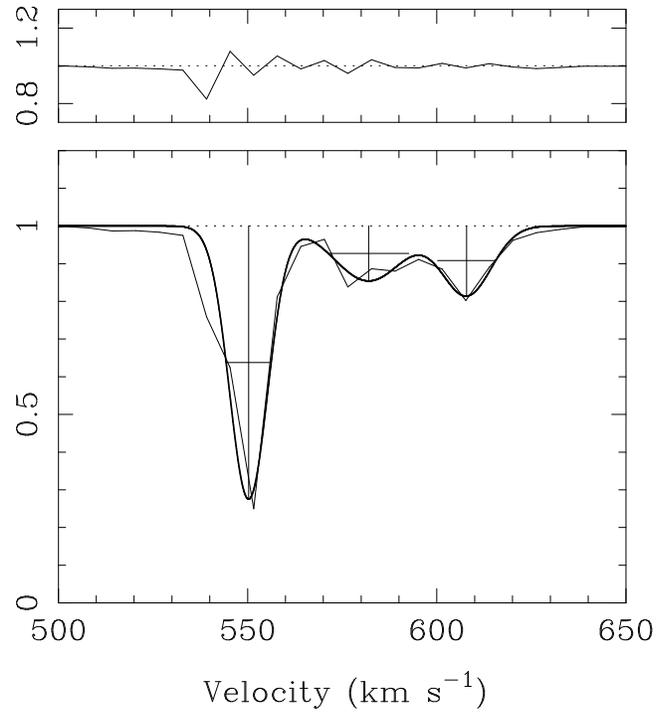}
\caption[]{The HCO$^+$(1--0) spectrum binned to a velocity
resolution of 6.2\,\kms, which is the same as the 21cm HI
spectrum obtained by van der Hulst et al. (1983). Also shown
is a 3--component gaussian fit and the residual spectrum.}
\end{figure}

\begin{table}
\begin{flushleft}
\caption[]{Column densities for HCO$^{+}$ and HI$^{a)}$.}
\scriptsize
\begin{tabular}{ccccc}
\hline
 & & & & \\
\multicolumn{2}{c}{Comp.$^{b)}$}                 &
\multicolumn{1}{c}{N(HCO$^+$)}                   &
\multicolumn{1}{c}{N(HI)}                        &
\multicolumn{1}{c}{N(HCO$^+$)/N(HI)}             \\
 & & & & \\
\multicolumn{1}{c}{V(HCO$^+$)}                   &
\multicolumn{1}{c}{V(HI)}                        &
\multicolumn{1}{c}{cm$^{-2}$}                    &
\multicolumn{1}{c}{cm$^{-2}$}                    &
\multicolumn{1}{c}{$10^{-8}$}                    \\
 & & & & \\
\hline
 & & & & \\
550 & 553 & 13.683 & 21.415 & 0.5 \\
582 & 576 & 13.098 & 21.279 & 1.5 \\
608 & 596 & 13.063 & 20.708 & 0.4 \\
 & & & & \\
\hline
\end{tabular}
\ \\
Column densites are given as $\log(N/{\rm cm}^{-2})$.\\
a)\ HCO$^+$ is smoothed to a velocity resolution of 6.2\,\kms. \\
Column densities for HI derived assuming $T_{\rm s}=100$\,K. \\
b)\ Center velocities for the HCO$^+$ and HI lines (see Fig.\,10).
\end{flushleft}
\end{table}

\subsection{Location of the absorbing molecular gas}

The presence of two absorption complexes in Cen A, one
at the systemic velocity and one redshifted relative to
the systemic velocity, has led to speculations that the
redshifted component arises in gas falling into the
nucleus of Cen A, possibly feeding a supermassive black
hole (cf. van der Hulst et al. 1983, van Gorkom et al.
1989). One argument is that the redshifted HI component
is only seen towards the radio core and not against
the inner jet (van der Hulst et al. 1983). The detection
of $\lambda$2cm H$_2$CO in absorption against the inner
jet (Seaquist \& Bell 1990) is not necessarily a proof
against the infall hypothesis, if the inner circumnuclear
molecular disk is extended on scales of $\sim$300\,pc.
On the other hand, the lack of redshifted HI absorption
against the inner jet could also mean that while the
size of the HI component at the systemic velocity is
$\ga$300\,pc (in order to cover both the jet and the core),
the extent of the redshifted gas is smaller.

HI absorption has been detected in 9 elliptical galaxies
(van Gorkom et al. 1989, Mirabel 1990 and references therein).
In all cases the absorption is redshifted with respect
to the systemic velocity.
It is not clear whether this preponderance of redshifted
absorption reflects a true infall of gas or if it results
from a systematic offset in the systemic velocities towards
the blue. Such systematic errors are known to exist due
to outflow of emission line gas, which is often used to
derive the systemic velocity.
In spiral galaxies the situation is different. Here HI
absorption is often seen as a single broad line, extending
into both the blue-- and redshifted sides around the systemic
velocity (cf. Dickey 1986).
With higher spatial resolution, the broad HI component
is decomposed into a narrow one, shifting across the
finite extent of the background radio source, over all
velocities defined by the rotation of the disk (Koribalski
et al. 1993). These absorptions originate in fast rotating
circumnuclear disks or rings, of size $\sim$200\,pc, and
with velocities $\sim$200\,\kms\ (e.g. NGC 253, 660, 1808,
3079, 4945, Milky Way).

\medskip

The same phenomenon could be occuring in the center of Cen A,
where the presence of a nuclear disk or ring of $\sim$100\,pc
radius and rotating with a velocity of $\sim$220\,\kms\ 
has been established (Rydbeck et al. 1993).
High rotational velocities at small galactocentric distances
are naturally occuring in elliptical galaxies due to the
high mass concentration towards the center (cf. Hernquist 1990)
and non--circular orbits for the gas can be generated through 
non-axisymmetric gravitational instabilities (e.g. bar) or
in the form of tri--axiality of the elliptical galaxy itself.
Since the angular extent of the continuum source in Cen A is
$\la$2\,mas (6000\,AU), either blue or redshifted gas,
depending on the orientation of the orbits, can be seen in
absorption.

This situation  is reminiscent of what happens in the center of
the Milky Way. Here the velocities are more strongly non--circular,
maybe because the central bar is oriented at 20--30$^\circ$ from
the Sun line of sight (e.g. Blitz \& Spergel 1991, Weinberg 1992). 
HCO$^+$ absorption in front of the central continuum source SgrA
reveals a broad component between $-$210 to $-$110\,kms, interpreted
as coming from the $\sim$200\,pc nuclear disk, and four narrower
features (at $-$51, $-$30, $-$2 and $+$32\,\kms) corresponding to
known spiral arms in the galactic disk (Linke et al. 1981).
The broad component at negative velocities is also seen in
absorption in front of SgrB2, in HCO$^+$ as well as H$^{13}$CO$^+$
and HCN (Linke et al. 1981, Greaves \& Nyman 1996).
The size of the millimeter continuum source SgrA has recently been
determined through VLBI at 3 and 7 mm (Krichbaum et al. 1994), and
is 0.33 and 0.75 mas respectively. At those frequencies, the
interstellar scattering becomes negligible, so these figures are
believed to be the actual source sizes (Krichbaum et al. 1994).
 Since these source sizes correspond to $\sim$5 AU at the Galactic
Center distance, the HCO$^+$ absorption features prove that
the apparent line of sight velocity dispersion can be quite high
in a typical edge--on  nuclear disk, due to the accumulation on
the line of sight of differential non--circular motions.

The emitting molecular gas in Cen A is confined to the center region,
consisting of the nuclear disk or ring at a radius of
$\sim$100\,pc and an outer ring (or spiral arm) at $\sim$750\,pc.
The absorption complexes are likely to be associated with these
features, but since absorption is sensitive to diffuse and low
excitation molecular gas which is generally not seen in emission
(e.g. Lucas \& Liszt 1996), it is possible that some intervening
molecular gas detected in absorption is associated with the larger
HI disk extending to 7\,kpc (Schiminovich et al 1994).
As we have seen in the previous analysis, the molecular gas in
both LV and HV components is diffuse, with abundance ratios similar to
Galactic values. The main difference is that gas in the
LV components has a higher kinetic temperature than the HV components.
We cannot differentiate between inner and outer molecular gas
from this alone.
There are, however, two facts which suggest that the HV components
are associated with the nuclear gas and the LV components with the
outer disk:
(1) the LV components are close to the systemic velocity and,
(2) whereas the LV components extend between 540--556\,\kms, the
HV components are spread out between 576--640\,\kms, $\sim$4 times
larger.
Although the rotational velocities are similar for the inner and
outer molecular gas (Rydbeck et al. 1993), the inner region has a
larger velocity gradient and this gas seen in absorption should
be spread over a larger velocity interval.

\medskip

The blue-shifted `wing', from 500 to 540\,\kms, has not been
considered above. This feature is seen at a very low level,
and depends on the subtracted emission profile shape, and
should be viewed with caution. 
Could this gas comes from inside the cavity delineated by
the circumnuclear ring? There is a constraint on the radius
where molecules can subsist, around a luminous ionizing X--ray
source. Maloney et al. (1994) derived an effective ionization
parameter $\xi_{eff}$:
$$
\xi_{eff} = 1.1 \times 10^{-2} L_{44}/(n_9 r_{pc}^2 N_{22}^{0.9})
$$
where $L_{44}$= L$_x/10^{44}$erg s$^{-1}$,
$n_9 =$ n(H$_2$)/10$^9$ cm$^{-3}$ and
$r_{\rm pc}$ is the distance from the X--ray source in parsecs.
The gas will not become substantially molecular unless the
effective ionization parameter is smaller than 10$^{-3}$.
From ROSAT HRI measurements, D\"obereiner et al (1996) have
estimated  an X-ray luminosity (0.1-2.4 kev) for the
Cen A nucleus of L$_{\rm x}= 3 \times 10^{41}$ erg s$^{-1}$
in 1994, after correcting for absorption.
For diffuse gas with n(H$_2$)$\sim 3 \times 10^{3}$\,\cmcb,
and a column density N(H$_2$) $\la$ 10$^{22}$ cm$^{-2}$,
the minimum distance from the center is $\sim$100\,pc.
For molecular gas to exist at smaller galactocentric distances
in Cen A, both the volume and column densities must be
considerably higher.
It is therefore likely that the molecular gas corresponding
to the blue--shifted wing is at least at the distance of the 
circumnuclear ring. It is possible that this wing corresponds
to the 750\,pc component, which should also possess non--circular
motions. The orientation of orbits in a tumbling non--axisymmetric
component change by 90$^\circ$ at each resonance (e.g.
Contopoulos \& Grosbol 1989). Along the nucleus line of sight,
it is therefore possible that this blue--shift component
corresponds to elliptical streamlines perpendicular to that
in the circumnuclear ring.

\section{Summary}

We have obtained new high signal--to--noise molecular absorption spectra
towards the Centaurus A radio core. This has allowed the identification 
of up to 17 components in the HCO$^+$ spectrum. 
The  components associated with the systemic velocity (LV) correspond to 
intervening material in the accreted gaseous disk of Cen A, well outside
the circumnuclear ring in non--circular motions. It is not
possible at present to derive their exact distances from the
center. The high velocity (HV) redshifted components are interpreted
in terms of a nuclear ring, at about 100\,pc distance, with
non--circular motions. The H$^{13}$CO$^+$ spectrum indicate
that only the central component has relatively moderate
opacities, while the HV components have low opacities. From
the comparison between the HCO$^+$, HCN, HNC and CS spectra,
we deduce that the absorbing gas is diffuse and cold on average.
Abundances are compatible with Galactic values.

Comparison with the HCO$^+$ spectrum obtained 7 years ago by
Eckart et al. (1990) revealed no time variations
at a level $>$10\%.
Changes were expected from similar Galactic experiments.
We therefore suggest that this constrains the apparent size of
the mm continuum source to be larger than 500 AU (or $\sim$0.2mas),
unless the absorbing gas is made up of a very large number of very
small (a few tens of AU) clumps. In the latter case variations
tend to average out over time.
The core at 8.4\,GHz appears as a point source ($\la$2mas) in VLBI
experiments (Jones et al. 1996). The source size is therefore
constrained between 0.2 and 2\,mas. The larger size might be only
apparent, enlarged through interstellar scattering by the ionised
gas in Cen A itself. 

\acknowledgements
We thank R. Lucas and H. Liszt for communicating Galactic column
density data prior to publication.
TW acknowledges support from the Swedish Natural Science Council (NFR)
for this research.


\end{document}